\documentclass[]{IEEEtran}

\IEEEoverridecommandlockouts

\input epsf
\usepackage{graphicx}
\usepackage{url}
\usepackage{amsmath}
\usepackage{xcolor}  
\usepackage{cite}

\begin{document}

\IEEEpubid{\makebox[\columnwidth]{979-8-3315-4683-0/26/\$31.00~\copyright2026 IEEE \hfill} \hspace{\columnsep}\makebox[\columnwidth]{ }}

\title{\LARGE Design and Verification of a New Current Source for \\Tsinghua Tabletop Kibble Balance}

\author{\IEEEauthorblockN{Yuhan Ma, Kang Ma, Wei Zhao, Lisha Peng, Songling Huang, Shisong Li$^\dagger$} \\
Department of Electrical Engineering, Tsinghua University, Beijing 100084, China\\
$^\dagger$Email: shisongli@tsinghua.edu.cn}

\maketitle

\IEEEpubidadjcol

\begin{abstract}
This paper presents an ultra-stable current source tailored for the one-mode, two-phase (OMTP) measurement scheme in Tsinghua tabletop Kibble balances. To achieve simultaneous high resolution and nA/A-level stability, a composite 'coarse-fine' control topology is proposed, utilizing a dual-DAC architecture and an active digital feedback loop. Experimental results show that the Allan deviation reaches 1\,nA/A at an integration time of approximately 3 minutes, representing a tenfold improvement in measurement speed compared to commercial-source-based setups. Furthermore, the design offers a significant cost advantage, providing a satisfying option for high-precision, cost-effective mass realization.
\end{abstract}


\pagenumbering{gobble}

\section{Introduction}

Following the 2019 redefinition of the International System of Units (SI), the Kibble balance has become a primary instrument for realizing the kilogram. Its operation relies on two measurement phases: weighing and velocity. For compact or tabletop Kibble balances, the OMTP scheme is often preferred to maintain thermal equilibrium \cite{Liu_Schlamminger_Li_2025, Li_Ma_Ma_Liu_Li_Liu_Peng_Zhao_Huang_Yu_2025, Li_Ma_Zhao_Huang_Yu_2023}. However, this approach imposes significantly more stringent requirements on the current source than traditional methods, demanding ultra-stable current during the velocity phase. As shown by the error propagation in the geometrical factor determined during velocity measurement, $\Delta(Bl) \approx I(\partial L/\partial z) + (L/v)(\partial I/\partial t)$ \cite{Li_Bielsa_Stock_Kiss_Fang_2017} (where $L$ is coil inductance, $I$ is coil current, and $v$ is coil velocity), any current drift directly affects $Bl$ and hence the mass measurement. To achieve a target mass uncertainty of $10^{-8}$, the current source must provide short-term stability of nA/A level and a resolution exceeding 26 bits—requirements that commercial general-purpose sources rarely satisfy simultaneously. A current source that merges two commercial sources is shown as a possible solution \cite{Ma_Liu_Zhao_Huang_Li_2024}; however, the stability time is slow ($\approx$30\,mins reaching 1\,nA/A) limited to GPIB communication delays and DVM readout. To address this gap, we have developed a custom precision current source employing a 'coarse-fine' topology specifically designed to meet the critical stability requirements of the OMTP scheme.

\section{Design}

\begin{figure*}[htbp]
    \centering
    \includegraphics[width=0.65\textwidth]{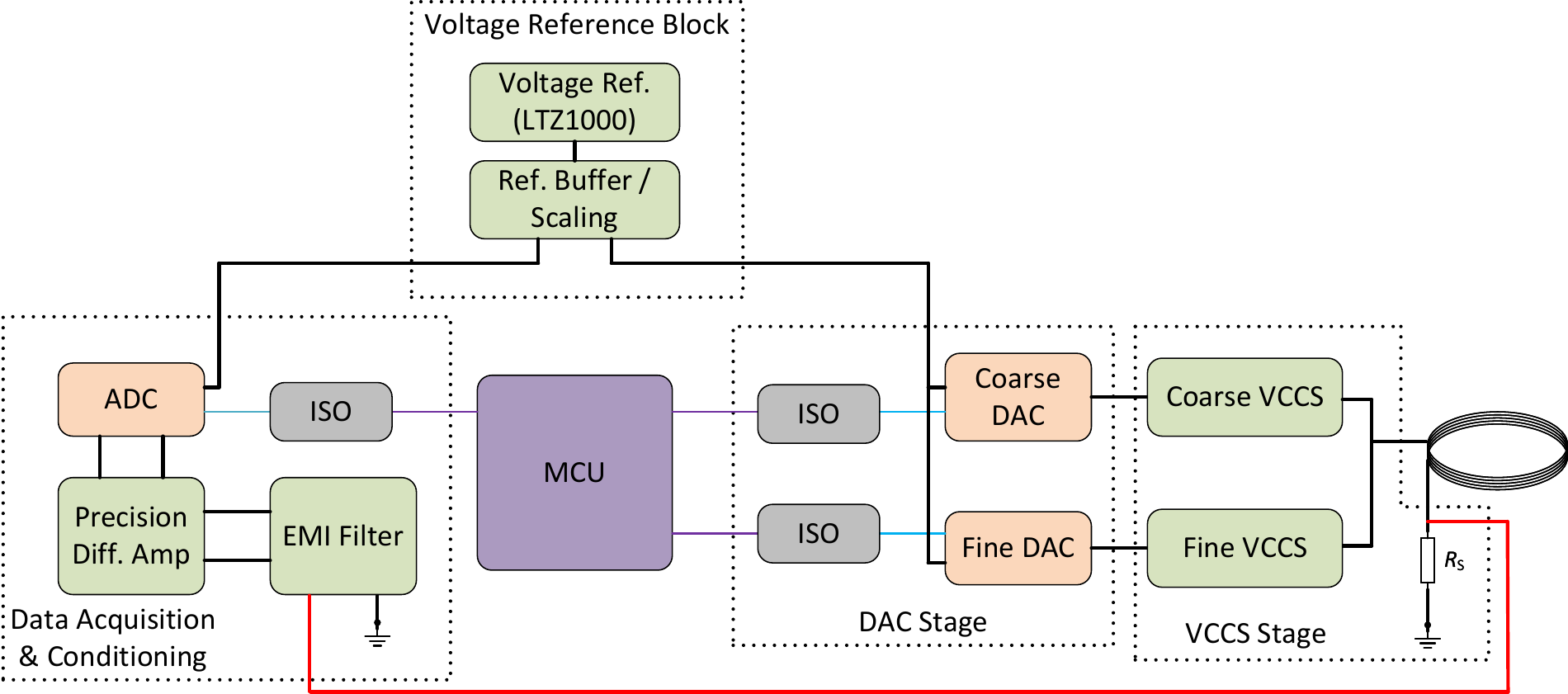}
    
    \caption{Simplified block diagram of the dual-path precision current source. The output is the sum of a coarse main current and a fine compensation current (attenuated by a factor of $K=200$ compared to the coarse).}
    \label{fig:overall2}
\end{figure*}

To meet the demanding combination of wide dynamic range (here $\pm 25\,\text{mA}$) and ultra-high stability ($10^{-9}$ level) required by the OMTP measurement scheme, this work proposes a composite 'coarse-fine' control topology. As illustrated in Fig.~\ref{fig:overall2}, the system splits current generation into two parallel paths: a main voltage-controlled current source (VCCS) providing coarse adjustment, and an auxiliary VCCS delivering fine compensation.
The total output current $I_{\text{out}}$ is the superposition of contributions from both paths. The fine path is attenuated by a factor of $K=200$ relative to the coarse path, significantly enhancing resolution. A 32-bit sigma-delta ADC monitors the voltage across a precision sensing resistor $R_s$, enabling a microcontroller (MCU) to close the feedback loop. The MCU computes the error between the setpoint and measured value, then adjusts the fine DAC in real time using a PID algorithm to actively suppress drift and noise. The resulting output relationship is given by:
\begin{equation}
I_{\text{out}} = \frac{V_{\text{DAC1}}}{R_{\text{V/I}}} + \frac{V_{\text{DAC2}}}{K \cdot R_{\text{V/I}}}
\label{eq:output}
\end{equation}
where $V_{\text{DAC1}}$ and $V_{\text{DAC2}}$ are the output voltages of the coarse and fine DACs, respectively, and $R_{\text{V/I}}$ denotes the transconductance gain of the VCCS stages.

The voltage reference is derived from an ultra-stable Zener diode, the LTZ1000, which exhibits a temperature drift of $0.05\,\text{ppm}/^\circ\text{C}$. Signal generation is handled by two 20-bit DACs. To minimize quantization error, the theoretical current resolution $I_{\text{LSB}}$ is determined by the fine control loop and is given by
\begin{equation}
I_{\text{LSB}} = \frac{V_{\text{ref}}}{2^{N}} \cdot \frac{1}{K \cdot R_{\text{V/I}}}.
\end{equation}
With $V_{\text{ref}}=10\,\text{V}$, $N=20$, and an attenuation factor $K=200$, the resulting resolution is approximately $238\,\text{pA}$. This realizes an equivalent dynamic range of 26.6 bits, thereby satisfying the $10^{-9}$-level resolution required for precise force-feedback control.
The sampling resistor $R_s$ is a critical determinant of overall current stability. To achieve high performance, a $200\,\Omega$ standard was constructed by connecting ten $2\,\text{k}\Omega$ metal foil resistors in parallel. This parallel configuration offers two key advantages: it distributes the thermal load to mitigate self-heating effects, and it leverages statistical averaging to achieve an effective temperature coefficient of $0.5\,\text{ppm}/^\circ\text{C}$ or better.

\begin{figure}[htbp]
    \centering
    \includegraphics[width=0.85\columnwidth]{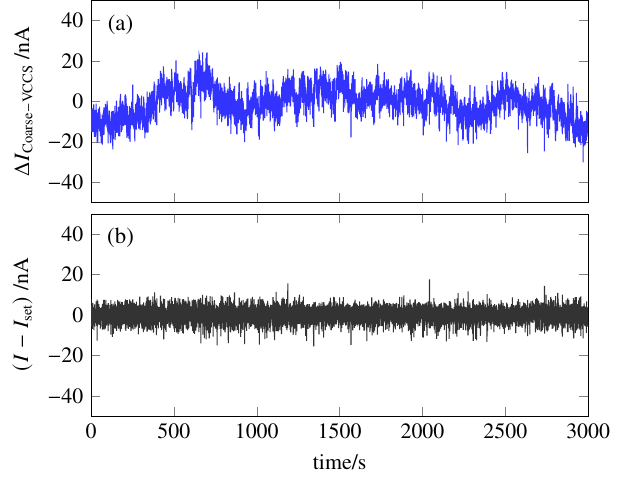}
    
    \caption{Time-domain current fluctuations for the 10 mA output. (a) shows the residual of the coarse main source relative to the setpoint ($I_{\text{Coarse}} - I_{\text{set}}$), showing inherent thermal drift. (b) presents the final total output residual ($I - I_{\text{set}}$) where $I_{\text{set}}$ is the set current value.}
    \label{fig:I_domain}
\end{figure}

\section{Experimental Verification}

\begin{figure}[htbp]
    \centering
    \includegraphics[width=0.85\columnwidth]{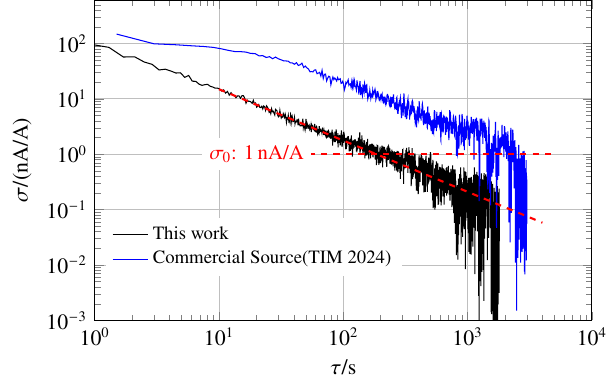}
    \caption{Comparison of stability performance between the proposed design and the previous commercial-source-based setup in \cite{Ma_Liu_Zhao_Huang_Li_2024}. The proposed design significantly reduces the integration time required to reach high stability.}
    \label{fig:allan_comparison}
\end{figure}

Stability verification was performed using data acquired from the internal 32-bit ADC. With an optimized PID controller and a 0.4\,s moving average filter, the system was evaluated in a constant-current mode, emulating operation with a target current of $10\,\text{mA}$.
Time-domain measurements reveal that the coarse main source alone exhibits a drift of approximately $40\,\text{nA}$ over a 50-minute measurement period. However, with active compensation from the fine control loop, the residual error of the total output current ($I - I_{\text{set}}$) is maintained within $\pm 10\,\text{nA}$.
To quantify stability, the Allan deviation $\sigma(\tau)$ was calculated. As shown in Fig.~\ref{fig:allan_comparison}, the stability follows a $\tau^{-1}$ slope, indicating effective white-noise averaging. At an integration time of $\tau \approx 100\,\text{s}$, the system achieves a stability of $1\,\text{nA/A}$ across the full dynamic range.

The proposed design was compared against a previous-generation setup based on commercial source meter units presented in \cite{Ma_Liu_Zhao_Huang_Li_2024}. The results demonstrate a substantial improvement: the integration time required to reach $1\,\text{nA/A}$ stability is reduced by a factor of 10 (from $30\,\text{mins}$ to $3\,\text{mins}$), and the best achievable stability level is improved from $>1\,\text{nA/A}$ to $0.3\,\text{nA/A}$.
These results confirm that the proposed current source satisfies the stringent stability requirements of the OMTP velocity phase in tabletop Kibble balance applications.

\section{Conclusion}
This work has presented an ultra-stable current source based on a dual-path VCCS architecture, specifically designed to satisfy the stringent requirements of the OMTP scheme implemented in the Tsinghua tabletop Kibble balance. 
Experimental validation confirms that the source reaches $1\,\text{nA/A}$ stability in three minutes of integration — a tenfold improvement in speed compared to setups relying on commercial source meter units. Moreover, the design achieves a reduction in cost by over 80\%, offering a robust and economical solution for Kibble balance measurement and other applications.

\section{ACKNOWLEDGMENT}
This work was supported by the National Natural Science Foundation of China under Grant 52377011.

\bibliographystyle{IEEEtran} 
\bibliography{ref}

\end{document}